# Implementing Decentralized Per-Partition Automatic Failover in Azure Cosmos DB

Josh Rowe, Mikael Horal, Hari Sudan Sundar, Muthukumaran Arumugam, Burak Kose, Sravani Mitra Palivela, Geni Marsh, Varun Jain, Abhishek Kumar, Dhaval Patel

Microsoft
One Microsoft Way
Redmond, WA 98052 USA

{joshua.rowe, mikaelhoral, hari.sudan, muthar, burakkose, srpalive, gferns, varunj, abk, dhavalpatel}@microsoft.com

## ABSTRACT

Azure Cosmos DB is a cloud-native distributed database, operating at a massive scale, powering Microsoft Cloud. Think 10s of millions of database partitions (replica-sets), 100+ PBs of data under management, 20M+ vCores. And failovers are an integral part of distributed databases to provide data availability during outages (partial or full regional outages). While failovers within a replica-set within a single region are well understood and commonly exercised, geo failovers in databases across regions are not as common and usually left as a disaster recovery scenario. An upcoming release of Azure Cosmos DB introduces a fine grained (partition-level) automatic failover solution for geo failovers that minimizes the Recovery Time Objective (RTO) and honors customer-chosen consistency level and Recovery Point Objective (RPO) at any scale. This is achieved thanks to a decentralized architecture which offers seamless horizontal scaling to allow us to handle outages ranging from node-level faults to full-scale regional outages. Our solution is designed to handle a broad spectrum of hardware and software faults, including node failures, crashes, power events and most network partitions, that span beyond the scope of a single fault domain or an availability zone.

In this paper, we present the design and implementation of "Per-Partition Automatic Failover". We introduce the "Failover Manager" which executes a deterministic, distributed state machine for each partition to coordinate and execute failover state transitions at partition-level granularity. We also introduce a novel CAS Paxos [1] implementation of a highly available and durable store which is used by the Failover Manager.

## 1 INTRODUCTION

Cosmos DB [2] is a large-scale distributed database service that supports multiple consistency levels [3] across geographic "regions". Horizontal scalability is achieved by organizing data into "partitions" [4] containing data from non-overlapping hash-based key ranges. Partitions are organized into "partition-sets" and "accounts". An "account" is a unit of management for users. A partition-set consists of a set of geographically distributed replicated copies of a single key range. Each partition resides in a single geographical "region". An account contains all the partition-sets that span the entire key range of all the databases present in an account. In a "single-writer" setup, each account has a single "write region" that serves read and write operations and zero or more "read regions" that serve only read operations. Within each partition-set, users issue read operations against partitions in "read regions" and issue read or write operations to a single "write region" partition.

Like all distributed systems, Cosmos DB faces failures from internal and external sources such as datacenter power failures, network partitioning, high resource consumption and software bugs. Some of these failures cause a subset of partitions within the current write region to fail or degrade. Prior to the work described in this paper, Cosmos DB handles such failures by attempting to "failover" an entire account to another geographic region, even if only a single partition has failed. This failover is signaled to clients via a DNS update and transparently handled by the SDK.

The process of failing over all partition-sets in an account can itself fail. For example, a read region partition within a large account might fail during the failover transition. This process is coordinated using a "control plane" set of machines, and the control plane itself can fail due to various reasons. The overall result is that the impact and risk of failing over an account increases with the size of the account and the overall scale of the service.

Geo-failover operations are typically triggered by an operator several minutes into an outage, and it can take time to failover all partitions of all affected accounts in a big fleet of accounts that a single customer could have deployed. Also, in our experience, the impact could start small and later spread in unanticipated ways. This could lead to delays in restoring availability, due to the operator judgement involved in deciding when to failover, and due to failing over their entire fleet of accounts, even if only a partial set of partitions were impacted. The combination of having an extraordinarily large failover impact for single partition outages, reaching the scaling limits of the control plane in the event of a broader outage, and the duration of availability loss during failures suggests a new approach should be considered for failovers.

The Cosmos DB team set out to implement a scheme internally referred to as "Per-Partition Automatic Failover". This scheme allows each individual partition-set to autonomously choose which region is the current write region, allowing independent decisions to be made for each partition-set. This reduces the number of partitions that need to perform work during partial outages that affect only a few partitions. During larger outages, the ability for the partition-sets to act autonomously reduces the risk that a control plane failure prevents successful restoration of availability of data plane for users. A heartbeat-based failure detection mechanism at the partition-set level automatically triggers failovers within a single partition-set. We target completing region-wide failovers within two minutes of failure being detected.

## 2 BACKGROUND ON COSMOS DB

A Cosmos DB partition is made up of a set of replicas within the region and is currently architected to choose a single long-term replica per partition (the "primary") whose responsibility is to accept write traffic, then to replicate that write traffic in a tree topology to other replicas. Consensus protocols such as Paxos [25], Multi-Paxos, Egalitarian Paxos [20], and Raft [23] do not strictly require a long-term leader; any leader can begin issuing new rounds of consensus in each of these algorithms. However, choosing a long-term leader typically allows accepting read and write traffic at much high performance. Cosmos DB follows this tradition of selecting a long-term leader.

Cosmos DB's consistency level offerings [21] and replication strategy evolved over time, leading to its current implementation. Cosmos DB arranges each partition into a set of replicas, each replica being in one of the roles "primary" or "secondary". These replicas, making up a partition, could be distributed across fault domains and availability zones, within a region. At any time, there is only a single primary replica. Typical operation uses 3 secondary replicas. Only a primary replica can accept "write traffic" from external sources. The primary replica transmits write operations to the secondary replicas in that partition and performs a quorum commit before acknowledging the user.

This arrangement supports the documented consistency levels of Cosmos DB for single-region accounts. Support for multi-region accounts was enabled by adding two new replica roles, "XP primary" and "XP secondary", where "XP" means "cross-partition". (These roles also support partition split and merge operations, not discussed at length in this paper.) The primary replica in each partition chooses an "XP primary" replica from the set of secondary replicas. For all read region partitions, the primary replica acts as an "XP secondary". The write region's "XP primary" replicates traffic it receives to the "XP secondary" replicas, which are the read region primary replicas. This architecture is visually depicted, albeit with different terminology, in the diagram from [16], "Replica Sets".

Figure 1 shows a single partition-set, where each "replica-set" represents a single partition. The "forwarder" is the XP primary. The "leader" replica in the left replica-set is the write region primary replica. The two other "leader" replicas represent "XP secondary" replicas. The remaining "followers" are secondary replicas. In this arrangement, a write operation is "committed" to a partition when a quorum of replicas in that partition has written the write operation to disk. A write operation is "globally committed" if a quorum of partitions in a partition-set have committed the write.

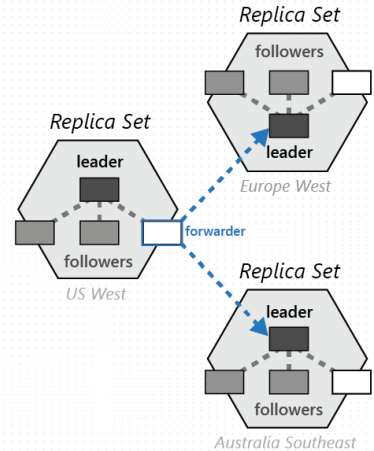

**Figure 1: Partition-set illustration. US West is the write region whose XP primary is responsible to replicating traffic to the read regions.**

Cosmos DB also has a multi-region-write offering. All regions are writable in this setup and hence there isn't a need to elect a new "write region" for the partition-set, since all partitions are writable.

So, Per-Partition Automatic Failover discussion is restricted to single-writer setup, where there is a user designated "write region" and other regions are in "follower/read-only" capacity.

Cosmos DB uses lease management to make consistency guarantees. This matters most in its "Global Strong Consistency" offering. [3] Cosmos DB is architected to serve read-only queries from its secondary replicas and read regions. Replicas that do not respond to replication messages have their leases revoked or expired and will fail to serve queries until they begin responding to replication messages.

Management of the current primary replica within a partition (a replica-set) is coordinated using Service Fabric. Service Fabric uses Paxos internally to perform state changes within a Service Fabric cluster of machines. Service Fabric has the capability of acting across geographic regions, and so in theory could coordinate Cosmos DB's failover schemes internally. [5] However, Cosmos DB consists of hundreds of separate Service Fabric clusters, and users can dynamically choose in which geographical regions their accounts are located. Service Fabric's notion of leader selection does not mesh well with Cosmos DB's geo-failover needs, so a new mechanism supporting Cosmos DB was needed.

## 3 PER-PARTITION AUTOMATIC FAILOVER

This section provides a summary of the design of Per-Partition Automatic Failover as well as the requirements which drove our design choices. We introduce the "Failover Manager" which implements the distributed state machine functionality for enabling horizontally scalable, decentralized per-partition failover.

## 3.1 Requirements

We began the process of implementing this feature by starting from these high-level requirements:

1) Cosmos DB must detect and recover from partition/regional failures in under 2 minutes.

    While any availability loss is undesirable for a customer, the core architecture of Cosmos DB is based on a lease management system. Using these schemes requires accepting the potential for availability losses on the order of the lease duration. This requirement is defined on a per-partition basis, but it is possible that many partitions may be simultaneously impacted. For example, a complete power loss to a data center may impact hundreds of thousands of partitions. The failover of these partitions must *all* complete within the desired time period.

2) The current set of account features must be supported.

    The feature set of Cosmos DB contains things such as: multiple consistency levels, change feed, backup and restore, dynamic capacity management, auto scaling, automatic indexing, adding and removing regions. These features must continue to work correctly in the presence of Per-Partition Automatic Failover. Notably, global strong consistency's defining characteristic is that there is never observable data loss for acknowledged write operations; other consistency levels permit some level of data loss for unrecoverable regional outages, but the advertised limits of allowable data loss must be respected.

3) The Per-Partition Automatic Failover feature must be compatible with the existing replication protocol, operational protocols and network topologies.

    Cosmos DB contains millions of lines of code. We deemed that reimplementing our replication protocols would be overly expensive and risky to the product. There are many operational protocols that Cosmos DB supports that must also continue to work. These operational protocols include partition-level migrations, splits, and merges, the ability to add and remove regions to and from database accounts, our internal configuration management systems, and our internal monitoring and troubleshooting systems. All these internal sub-systems have assumed a singular write region at account level so far; going forward, they will have to align with partition level write region going forward.

4) Customer latency requirements should be respected.

    Cross-region request processing latency is higher than intra-region request processing latency due to network transit times. Customers typically organize their Cosmos DB accounts so that write traffic is transmitted within a single region, from the customer's application hosted in Azure to the Cosmos DB partitions in that region. Customers usually prefer latency losses over availability losses, but when lower latency is possible, it should be prioritized.

5) The feature must be transparent to users *not* using direct mode [24], but can require client SDK updates by customers using direct mode.

    The Cosmos DB SDK has two modes: direct mode and gateway mode. Direct mode creates TCP connections directly to the backend hosts performing data operations. Gateway mode connects to the frontend gateway machine, which then uses the Cosmos DB SDK in direct mode. New features that affect the network design of the Cosmos DB SDK require customers to begin using that new version of the SDK. Other application-level changes might also be required, but it is desirable to limit these.

## 3.2 High-level design

We chose a design that would be minimally invasive to Cosmos DB, identifying integration points where the feature would operate. We defined a state machine that drives behavior changes. This state machine is backed by a separate highly available store using the CAS Paxos protocol to coordinate updates. State machine changes trigger existing behaviors in the backend processes. We call the backend component that implements this behavior the "Failover Manager".

Replicas in each partition perform state updates to transition a partition-set from state to state. Each state update performs a CAS Paxos round as a leader, using the state machine transition function as the edit function to apply. The result of updating the state machine is then translated into actions for that replica to apply to its local runtime state. Example actions are:

- To begin acting as a write region primary replica.
- To begin acting as a read region XP secondary replica.
- To stop accepting new write traffic in preparation for a graceful failover.

## 4 THE FAILOVER MANAGER STATE MACHINE

At the core of Per-Partition Automatic Failover is the new "Failover Manager", which executes a deterministic state machine for each partition to coordinate and execute failover state transitions at partition-level granularity.

This section discusses the Failover Manager state machine approach and discusses the implementation of a novel, highly available store based on CAS Paxos that the Failover Manager uses to store state machine state.

## 4.1 State machine vs. workflow approach

Using a state machine approach for geo-failovers is a change for Cosmos DB. Cosmos DB currently uses an internal control plane to manage accounts, partitions, and partition-sets. This control plane executes "workflows" that coordinate backend operations to effect state changes. The Cosmos DB Control Plane, that orchestrates account-level geo failover, is built upon workflows that execute sequences of operations that each contain internal persisted state. In contrast, a state machine approach defines a set of persisted state data and formally specifies the allowable state changes. These workflows impose limits on Cosmos DB's release process, runtime reliability, and scalability.

Persisted workflow state is generally not backward and forward compatible across software releases. This is a problem that systems such as Windows Workflow Foundation attempt to address [6], but those solutions are quite complex to manage and sometimes do not address all possible changes: Windows Workflow Foundation does not attempt to migrate the state associated with running actions. The state machine's formally specified data allows us to guarantee that state information is forward and backward compatible between software versions, allowing software releases during state transitions.

Workflows must reach terminal states, either by success or failure. When a terminal state is reached via failure, the system may be in an arbitrary state, which then requires logic in the next workflow execution to recover from these arbitrary states. This methodology is prone to error and leads to availability losses for customers. The formally specified state machine has no terminal states, ensuring that availability is always eventually restored.

The Cosmos DB Control Plane is a relatively small number of physical machines compared to the backend clusters that host the data partitions. The control plane faces scalability limits when large numbers of accounts need to be failed over. Driving state machine changes directly on the backend machines allow scaling the volume of control operations directly with the backend machine count, removing the scaling limits of the control plane.

These workflows can be limited in their capabilities in the face of multiple simultaneous operations. Each workflow typically obtains a lock on its subject partitions, this lock being released near the end of the workflow. While the workflow is executing, no other workflow can acquire the lock. This prevents, for example, a failover workflow from executing at the same time as a partition split workflow on the same set of partitions. The state machine approach allows more arbitrary types of state changes, eliminating these simultaneity restrictions.

## 4.2 Distributed state machine execution

We desired that the Failover Manager component be able to make state transitions in a distributed fashion using the machines the partitions are hosted on to make decisions. This desire was driven by requirements of resilience and simplicity. Deploying a new service, with new RPC schemes, entails increasing the exposure to new failure modes. Making such a service scalable entails scaling the service along with the size of Cosmos DB. Making the service resilient to entire region failures means having multiple instances of the service; this service then needs to be able to coordinate its own state updates somehow, therefore requiring distributed protocols for executing state transitions.

Instead, we decided that the distributed protocol for executing state transitions would live directly in the backend service to be highly scalable with the size of the backend partition fleet. Each state transition accepts an input from the partition performing the state transition and performs a change to the persisted state. The newly persisted state can then be translated into a set of local actions to take.

The change to the persisted state is executed as a compare-and-swap operation. Each state change performs this algorithm:

1) Compute a "report" with the local status of the partition.
2) Read the current persisted state machine value and its version number.
3) Perform an edit operation using the state machine value and the report value as inputs and produce a new state machine value.
4) Perform a compare-and-swap operation on the persisted state machine value with the new state machine value, using the version read in Step 2 as the comparison value.
    a. If the compare-and-swap fails, go to Step 2.

The Failover Manager component executes infrequently: each primary replica in each partition performs a state update every 30 seconds. The probability of a conflict during the compare-and-swap operation in Step 4 is relatively low (but very much non-zero, which we discuss in the Experimental Results section).

## 4.3 CAS Paxos

The Failover Manager state machine's state must be persisted somewhere. This state persistence store must be highly available, consistent, resilient to failures, and be able to store a mutating value over time. We settled on CAS Paxos [1] as the ideal mechanism to provide availability, consistency, and mutability. The title of the CAS Paxos paper describes our reasons for this choice very succinctly: "CASPaxos: Replicated State Machines without logs". We desired a single replicated state machine for each partition-set, replicated across a globally distributed set of locations, where each update operation could be guaranteed to be considered by all future update operations, with minimal extra state management.

### 4.3.1 State machines

We implemented CAS Paxos in a multi-layer approach. The lowest layer transliterates the state machine operations described in the TLA+ [11] descriptions of Paxos [7] and CAS Paxos [8]. This layer contains classes for the roles defined in Paxos.

```
class LeaderStateMachine
{
    // The resulting Phase1aMessage should be sent to all
    // acceptors.
    StartPhase1Result StartPhase1(
        const Nullable<NakMessage>& message = {});

    // For each Phase1bMessage from an acceptor, call this
    // method. If the result is empty, then keep receiving
    // messages from acceptors. If the result contains a
    // Phase2aMessage, send it to all acceptors and begin
    // trying to learn the result from the resulting
    // Phase2bMessages and a LearnerStateMachine.
    template<typename TValueEditor>
    StartPhase2Result StartPhase2(
        const Phase1bMessage& message,
        TValueEditor valueEditor);
};
```

**Figure 2: CAS Paxos Leader state machine class definition**

```
class AcceptorStateMachine
{
    AcceptorStateMachine(
        AcceptorState acceptorState);

    // Upon receipt of a Phase1aMessage:
    // Call this method to update the acceptor's state,
    // Persist the state, then send the resulting
    // Phase1bResult back to the leader.
    Phase1bResult OnReceivedPhase1a(
        const Phase1aMessage& message);

    // Upon receipt of a Phase2aMessage:
    // Call this method to update the acceptor's state,
    // Persist the state, then send the resulting
    // Phase2bResult back to
    // the learners.
    Phase2bResult OnReceivedPhase2a(
        const Phase2aMessage& message);

    const AcceptorState& GetAcceptorState();
};
```

**Figure 3: CAS Paxos Acceptor state machine class definition**

```
class LearnerStateMachine
{
public:
    LearnerStateMachine(
        TQuorumCheckerFactory quorumCheckerFactory,
        LearnerState learnerState);

    // Upon receipt of a Phase2bMessage, attempt to
    // learn from it.
    // The LearnResult may be empty if no value is learned,
    // or a valid value if a value is stably learned.
    LearnResult Learn(
        const Phase2bMessage& message);

    const LearnerState& GetLearnerState() const;
};
```

**Figure 4 CAS Paxos Learner state machine class definition**

Writing this layer in this way ensured that we made no errors in translation; indeed, we've found no bugs in our CASPaxos state machine implementation.

The second layer implements message transmission and acceptor state storage using our application-level logic. This layer performs all three roles (Leader, Acceptor, and Learner) inside a single process, using external storage to persist the serialized acceptor state. Races to update the acceptor state storage are resolved by performing acceptor state machine changes using a compare-and-swap algorithm similar to that used for the Failover Manager state machine: failure to perform the compare and swap causes a re-read of the acceptor state, a re-application of the acceptor state machine to the message and state, and a retry of the compare-and-swap operation.

### 4.3.1 Choice of storage for acceptor state

We explored several options to store the CAS Paxos acceptor state. The requirements for this store are:

1) The store must be at least as low in the Azure stack as Cosmos DB.
2) The store must support a compare-and-swap operation on complex document content.
3) The store must be geographically distributed.
4) The store must be able to scale up to the workload imposed by this feature executing across all partitions in Cosmos DB during a regional outage.

Coincidentally, Microsoft Azure has an offering that meets these requirements called "Cosmos DB". We configure a set of geographically distributed *non-replicated* Cosmos DB accounts to act as the acceptor state to back all partitions globally. We use the 'If-Match' HTTP header to perform acceptor state updates atomically [9]. Cosmos DB implements horizontal scaling, allowing the acceptor state storage to scale to support all replicated partition-sets in Cosmos DB. This might appear to create a circular dependency from Cosmos DB onto Cosmos DB. This is illusory, as we are only creating a dependency from the Cosmos DB cross-partition replication stack onto a non-cross-partition replicated Cosmos DB account. Our choice of the actual storage provider is flexible enough that if this decision needs to be revisited, we can do so with relative ease.

We explored using a dynamically managed quorum using the data partitions themselves to back the acceptor state. This leads to complexities performing management of quorum sets in the face of adding and removing regions. This problem is solved and is described in [17][18][19]. One problem left unsolved in that approach is the 2-region scenario: the most inexpensive solution for customers to maintain high availability in the face of a region outage is to have two regions in an account. Using solely the partitions backing a 2-region account would disallow state changes in the case one region fails. It is possible to extend a quorum using a single extra external store. Windows Failover Cluster has a Cloud Store witness feature [10]. Rather than burden ourselves with these complexities in the initial implementation, we decided to begin implementation using a set of statically configured stores for acceptor state.

## 4.4 Failover Manager state machine

We specified the Failover Manager state machine using TLA+ and verified it using TLC, the TLA+ checker [11]. This allowed us to explore the behavior of the state machine in millions of scenarios in just a few minutes and allowed us to verify with high probability some properties of the state machine and its interaction with Cosmos DB consistency levels. Translating the state machine into C++ was then a trivial exercise.

The properties we desired of the state machine were:

- Absent the presence of repeated failures and failures greater than that required for maintaining quorums for each consistency level, it must always be true that a partition becomes available for read and write within a constant multiple of the replication lease expiration interval.

- When data is available to be read or written by a partition in a partition-set, the desired consistency level is maintained.

We did not have to apply the real-time methodologies described by [22 Ch. 9 "Real Time"], as we treated time as discrete quantity that ticks once per replication lease interval.

To avoid lengthy TLC verification using temporal properties, we used a scheme where certain variables were recorded in a special state history variable. We then verified invariants such as the 'WritesEnabledAtEndOfHistoryWhenRegionsSetIsStable' and 'ReadProperty' as seen in Figure 5.

```
WritesEnabledAtEndOfHistoryWhenRegionsSetIsStable ==
    /\ IF IsEnoughHistory(RegionStateHistory)
        THEN
          LET
            lengthOfFullHistory == Len(RegionStateHistory)
            lastRecentHistoryEntry ==
RegionStateHistory[lengthOfFullHistory]
            recentHistoryEntries == SubSeq(
              RegionStateHistory,
              lengthOfFullHistory - NumberOfHistoryTicksToLookback + 1,
              lengthOfFullHistory)
          IN
            IF
              /\ IsRegionSetStableInHistory(recentHistoryEntries)
              /\
IsUserPreferredWriteRegionStableInHistory(recentHistoryEntries)
            THEN
              /\ IsCurrentWriteRegionTakingClientWrites
            ELSE
              TRUE
        ELSE
          TRUE

ReadDataOnRegion(region) ==
  /\
     \/ RegionCurrentServiceStatus[region] = ReadOnlyReplicationAllowed
     \/ RegionCurrentServiceStatus[region] = ReadOnlyReplicationDisallowed
     \/ RegionCurrentServiceStatus[region] = ReadWrite
     \/ RegionCurrentServiceStatus[region] = ReadWriteWithWritesQuiesced
  /\ RegionCurrentBuildStatus[region] = BuildCompleted
  /\ RegionLatestCommittedUptoLSN[region] = RegionGCLSN[region]
  /\ LastReadData' = [Region |-> region, Data |->
RegionCommittedDataAtLSN[region][RegionLatestCommittedUptoLSN[region]]]

ReadProperty ==
  [][
    LastReadData'.Data >= LastReadData.Data
  ]_<< LastReadData >>
```

**Figure 5: Example TLA+ Properties**

## 4.5 Failover modes – Graceful and Ungraceful

We implement two failover modes: graceful and ungraceful. Each partition periodically updates the Failover Manager state machine. When the current write region partition continually fails to do so, an ungraceful failover is triggered. The Failover Manager state machine waits for a defined quorum of partitions to report state after the determination to perform an ungraceful failover has occurred, then chooses a new write region based on a user-defined priority list and based on examining the highest reported progress of all regions reporting; the highest priority region that shares the highest progress is then chosen. At any time, any region out of a defined quorum that is providing state updates and has the highest progress in that quorum can be selected. In consistency models weaker than "Global Strong" consistency, this can result in data loss; this data loss is accepted by the customer having chosen a weaker consistency model. We attempt to minimize this data loss by choosing the region with the highest progress after waiting a short period of time for regions to report their progress.

This model, instead of a workflow driven model, ensures that there is always a region available to fail over to in case the selected region also fails. This property was verified by TLC.

It is generally desirable for customers to access data in the region as close to the customer's application as possible. Regions are arranged by users in priority order. When a higher priority region becomes available to become the write region, the Failover Manager state machine begins performing a graceful failover to that region. A graceful failover suspends accepting writes for a short period of time, waits for all traffic to finish replicating to the new write region, then enables writes at that new write region. Users can change their priority list at any time; whenever there is a mismatch between the priority list and the current state of a partition-set, the Per-Partition Automatic Failover feature performs a graceful failover.

The process of a graceful failover itself can fail, either because the source or destination partition fails. When this happens, we simply initiate a new ungraceful failover. We detect this by a simple check: if too much time has passed while a graceful failover is ongoing, we perform an ungraceful failover. The state machine encodes this behavior.

There is a potential for degenerate behavior: it is possible for the graceful failover target to become responsive again, to be chosen as a graceful failover target, and for this graceful failover to again fail, leading to a loop where graceful failovers are continually attempted. A simple exponential backoff strategy on graceful failovers solves this problem. The count of the number of unsuccessful graceful failovers is stored in the state machine, along with the last time one was attempted. Graceful failovers are disallowed until an appropriate time. Without this fix, the degenerate behavior will cause a continuous outage. With this fix, the degenerate behavior affects the customer with increasing rarity; the customer will see outages, but they will rapidly decrease in frequency due to the exponential backoff.

There is another potential degenerate behavior: in a loop, a graceful failover can succeed, then destination region fails, and an ungraceful failover happens. We will amend our implementation to account for this by requiring exponentially increasing amounts of "live" time for a graceful failover target.

## 4.6 Dynamic Quorum

Historically, Cosmos DB implemented Global Strong consistency using a typical strict majority quorum scheme. A strict majority of regions would be required to acknowledge a write operation before the write operation could be acknowledged to the user. To enforce read consistency, all regions with an active "read-lease" must acknowledge that a write operation has taken place, even if the region has not yet committed the write. If a region does not respond, eventually its read-lease is terminated.

This scheme is problematic with two regions: the only strict majority in 2 regions is both regions, implying that availability is lost if either region fails. Even three region cases present issues: in the presence of a regional outage, it is possible for a single partition in a second region to fail which would lead to loss of write availability for that partition-set.

In our experience, customers typically prioritize availability in such situations and would prefer to maintain availability when there is only one available copy of their data. They do this with the expectation that yet another cascading failure is rare and expect that, upon recovery, they will eventually be able to make use of all replicas again, maintaining strong consistency in the meantime.

We therefore desired a protocol where any number of partitions in a partition-set can fail, gracefully degrading the number of active partitions. Prior art for this exists in e.g. Windows Failover Cluster [12]. We record the current set of read-leases in the Failover Manager State. When a partition must have its read-lease revoked, we first consult the Failover Manager and request permission. Permission is denied if the number of remaining read-leases (including the implicit write region's lease) would decrease below the user's configured minimum durability. Upon a failover, any partition that had an active read-lease can be chosen as the failover target.

With this scheme, a user can configure a two-region account with minimum durability 1. If either region's partition fails, the remaining active region's partition can take over, remove the failed region's partition from the set of active read-leases, and continue operating. When replication resumes and the previously failed partition begins acknowledging write operations, it can be re-added to the set of active read-leases, then re-added to the Failover Manager state, and it again becomes a potential failover target again.

## 5 INTEGRATION AND DEPENDENCIES

The backend partition is only part of the failover story. The capability of the backend service to fail over is irrelevant if customers cannot successfully use the system during and after such an event. Cosmos DB internally must also manage partition-sets when they are in failed-over states.

Our guiding principle here is that user-visible behavior must work correctly and transparently to users, but background system maintenance operations are permitted to be impacted until normal behavior is restored.

### 5.1 Client failover and SDK integration

The account-level failover features that Cosmos DB offers today are detected by the Cosmos DB SDK through DNS updates. The Cosmos DB SDK periodically re-resolves the DNS endpoint for the current write region. An account-level failover updates this DNS entry during the workflow that migrates the current write region for all partition-sets in the account to another region. This model has several weaknesses.

Updating a DNS entry requires that the DNS Time-To-Live field be respected at all intermediate DNS resolvers. Any customer, ISP, etc., who has updated their DNS client or intermediate DNS resolver to ignore the TTL field can cause DNS resolution to continue to return the old value [13].

Even before Per-Partition Automatic Failover existed, it is seen that during an account-level failover each partition-set is in different states at different times. During an account level failover, each partition-set transitions the current write region relatively independently from each other, though a trigger from the Control Plane. A client attempting to write by using the "current" cached write region for the *account* will find that certain migrated partition-sets will not accept writes. The client can attempt other regions, but prior to the Per- Partition Automatic Failover, the implementation did not have a cache of the current write region on a per-partition-set basis.

Finally, updating DNS records when there is already an issue at hand introduces more moving parts into restoring availability. Entire region failures might also cause the ability to write to Azure DNS to fail. Relying on more services to update data increases the surface area for failures.

We decided to use a single DNS TXT record per account to store each of the account's regional endpoints and their priorities. This information is written during account provisioning, when users add or remove regions to the account, or when users change region priority settings. During failover situations, no DNS updates are performed. The client detects failures to perform operations and attempts using other region's partitions as specified by the TXT record.

This new model exposed a flaw in error handling in the client. Previously, certain errors were not deemed to be retriable because (absent a DNS update) the request would be certain to fail. With Per-Partition Automatic Failover, these errors must always be interpreted to mean that the current write region for the partition-set is unavailable, and other regions should be tried: no DNS update will be forthcoming, and so the only evidence the client SDK can use to decide to try other regions is the error having been returned. Absent other evidence, every error becomes evidence of the need to try other regions. This evidence is collected into a per-partition-set cache, and regions are tried in order of the likelihood of success.

### 5.2 Control Plane and Capacity Management integration

The Cosmos DB Control Plane controls the horizontal scale and physical placement of partitions on hardware. Typical operations performed by the control plane include splitting partitions, migrating partitions, and adding a region to an account. The control plane also manages "Disaster Recovery" (DR) which is the process of reacting and responding to large-scale outages by forcibly failing over the write-region for every partition-set in one or more accounts away from an unhealthy region to a healthy region (during a large outage there may be 1000s or even 10,000s of accounts impacted). What these operations have in common is that they are all executed by the Cosmos DB Control Plane using account-level or partition-level locks. The various control plane workflows are written with the implicit and sometimes explicit assumption that they are being executed in exclusivity.

With Per-Partition Automatic Failover, this assumption no longer holds true as failovers are now initiated and executed in a decentralized fashion, without any synchronization or involvement from the control plane. While we *could* have decided to integrate Per-Partition Automatic Failover with the control plane for centralized management, this would have been contrary to our goals of horizontal scaling and high availability. The significant investment we have in the existing control plane operations prevented us from doing a large-scale refactoring of the existing workflows.

Through carefully reviewing most of the control plane workflows we identified the "account topology" to be the resource that required synchronization. The topology resource contains the current write-region for an account as well as the Global Configuration Number (GCN) which is the high-order bits for the partition-set's epoch which is incremented during failovers as well as for certain control plane operations. Per-Partition Automatic Failover introduced four challenges:

1) There is no longer the concept of an account-wide write-region in the topology; each partition-set's topology, at any point in time, may have a different write-region.
2) Changes to the topology are no longer protected by an account-wide lock.
3) The Failover Manager State is now the source of truth for the GCN and current write-region, so any changes to the topology need to be reconciled against this state.
4) Control plane metadata updates lost in the event of a concurrent failover, leading to inconsistent state between control plane and data plane.

The approach we took to fix issues (1) and (2) was to rely on optimistic concurrency control; any update to the topology is required to do a compare and swap operation. Failure by the control plane to perform an update would require a retry or alternatively rollback and cancellation. To fix (3) we introduced the concept of an "topology upsert intent"; rather than directly modifying the partition topology, a control plane workflow expresses an "intent" to do so, e.g. to revoke write-status for a partition, which the Failover Manager attempts to carry out by executing a full CAS Paxos round to update the Failover Manager State according to the desired intent. The control plane workflow in turn monitors whether the intent gets honored and needs to act accordingly, by either completing the operation, retrying, or canceling with rollback. (4) was addressed by introducing strong consistency semantics for metadata writes and reads; this ensures that an acknowledged write from the control plane gets persisted across failovers.

This approach only imposed limited additional complexity on the existing workflows since they already needed to handle cancellation and rollback scenarios. With Per-Partition Automatic Failover, we introduced the possibility that e.g. a partition migration operation would need to be cancelled and rolled back in case a failover for the partition happened while the migration was in progress. However, it is very likely that the partition migration would have had to be rolled back anyways (most likely due to a timeout) as the reason for the failover in the first place was due to the partition being "unhealthy".

## 5.3 Replication

One of our key design principles and goals was not having to re-write the Cosmos DB core replication protocol; largely we wanted to treat replication as a "black box". Our core replication protocol handles both in-region replication as well as cross-region replication and a rewrite would have increased both the scope and the risk of the project. With a few exceptions, we were able to accomplish this goal. This section shares insight into a few issues we ran into.

### 5.3.1 Partition reuse

Prior to Per-Partition Automatic Failover, bringing a partition back up following an outage entailed a full "reseed", i.e. wiping all the data from the partition and copying all of the data from the current write-region partition. This could take hours depending on the amount of data in each partition-set.

With Per-Partition Automatic Failover, our target was to reduce the failback duration to seconds or minutes (approximately proportional to the length of the outage). To accomplish this, we needed to be able to reliably determine the logical sequence number (LSN) in the old write-region up to which we needed to retain the data; any LSNs above this number would either need to be discarded or reconciled depending on consistency level and user preferences. We refer to this data as "false progress". To address this issue, we had to extend the replication protocol with a new dedicated "progress table" which tracks the LSNs written in each epoch. Using the progress table allowed us to undo any false progress as part of the failback process; this is needed to ensure consistency across partitions in the partition-set. It also enables us to only copy the delta of writes written to the new write-region during the duration of the outage.

### 5.3.2 Decentralized Cross Regional Leadership Election

Account level Geo-Failovers used to be coordinated by our centralized Cosmos DB Control Plane. This was a scaling bottleneck and a single point-of-failure. On the plus side, the control plane had a "global view" of each partition-set; by having the control plane assign and revoke leadership to a given region we were able to reduce the likelihood of "split brain" scenarios where more than one partition believes itself to be the write-partition.

By moving to a decentralized, distributed state-machine-based leader election, we introduced more scenarios where - e.g. during a network partition - multiple partitions may think they are the leader. Additionally, with Per-Partition Automatic Failover, partition-level failovers are no longer limited to infrequent large outages. In combination, this exposed previously unknown issues in our replication protocol.

Except for a couple of cases where we needed to make intrusive changes to the existing core replication protocol, we were able to leverage the Failover Manager and its Failover Manager State as the source of truth when it comes to cross-regional, geo leadership. This allowed us to accomplish our goals of minimizing changes to the existing core replication protocol.

# 6 EXPERIMENTAL RESULTS

## 6.1 Partition Failover

In this section, we go over the results from an exercise where we performed three power outages in a region hosting 4,300+ write-region partitions. Each power outage lasted for thirty minutes. The results demonstrate Per-Partition Automatic Failover's ability to restore write-availability within our Recovery Time Objective goal of two minutes. The results also demonstrate how, following power being restored, we are able to quickly failback to the original, preferred write-region.

### 6.1.1 Account Failover setup

To demonstrate the results from the experiment, we set up a 3-region account with the topology according to Table 1.

**Table 1: Account topology**

| Region Name | Status |
|---|---|
| East Asia | Write region |
| Southeast Asia | Read region |
| South Central US | Read region |

Table 2 lists the timestamps when each power outage was initiated. For each power outage, we brought down all the data plane nodes in East Asia for 30 minutes until we restored power again.

**Table 2: Power Outage Simulation Timestamps**

| Power Outage Sequence | Timestamps |
|---|---|
| Outage Simulation #1 | 2025-03-14T00:00:28.7029696Z |
| Outage Simulation #2 | 2025-03-14T01:00:47.1734173Z |
| Outage Simulation #3 | 2025-03-14T02:01:11.6935665Z |

### 6.1.2 Account Success Rates

Figure 6 shows how write-availability is maintained as the write-region for all partition-sets switch from East Asia to Southeast Asia across the experiment.

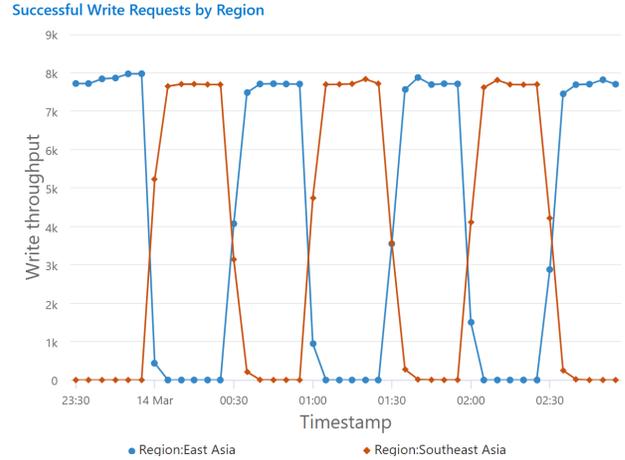

**Figure 6: Write throughput persists amidst power outages**

### 6.1.3 Partition Recovery Statistics

Across the three power outages, we recorded availability recovery time statistics for all partition-sets. Figure 7 demonstrates that availability is restored within less than 2 minutes for every partition-set, with the majority of the partition-sets showing recovery within a minute. Note that the Cosmos DB SDK implements client-side retries, so the customer measured availability loss is significantly less than the service-side metrics recorded in this graph imply.

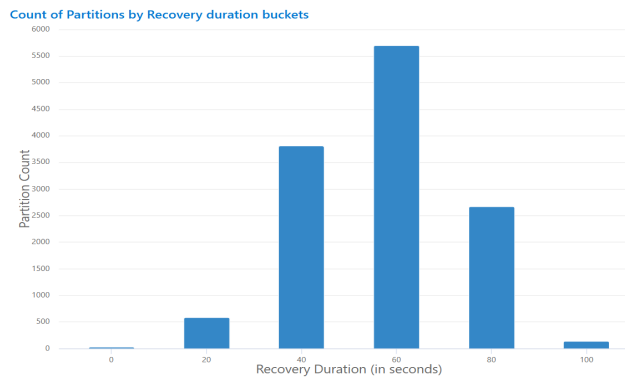

**Figure 7: Partition availability restoration time**

### 6.1.1 Outage recovery detection durations

Figure 8: Recovery detection time captures the time taken across all partition-sets to automatically detect recovery from a power outage in the preferred write region (East Asia), as partitions reach a healthy state for initiation of catch-up and graceful failover process. As seen below, the majority of the partition-sets detect this in 1 minute or less.

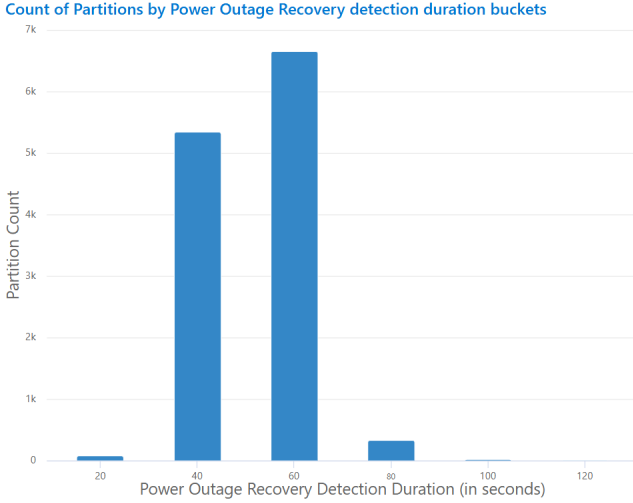

**Figure 8: Recovery detection time**

## 6.2 CAS Paxos

A key challenge in implementing decentralized Per-Partition Automatic Failover using CAS Paxos in Cosmos DB is the occurrence of dueling proposers. This phenomenon arises when multiple partitions in a partition-set simultaneously initiate CAS Paxos rounds, leading to conflicts that hinder efficient consensus.

Although dueling proposers is a well-known issue in consensus literature [14], its resolution typically involves designating a single distinguished proposer to serialize state updates. CAS Paxos, however, is inherently leaderless; any partition in a partition-set may propose state updates. Introducing a single distinguished proposer would undermine the decentralized design that is central to autonomous failover and reduced system dependencies.

To preserve CAS Paxos's leaderless property, we investigated more effective ways to handle NAK (negative acknowledgment) messages. A common industry practice is to employ exponential backoff with jitter, where the retry delay grows exponentially with each attempt, and random jitter offsets synchronous retry collisions. This approach is often summarized by a formula of the form:

$$\tau_{NAK} = \delta \cdot RandomUniform(0, 2^{(attempt-1)}) \quad (1)$$

where $attempt$ is the retry attempt count and $\delta$ is the base delay. While effective in many distributed systems, selecting a suitable base delay in heterogeneous network environments can be problematic. For instance, round-trip latencies between a user region in West US and an acceptor store in East Asia may reach a P50 latency of 150 ms (see [15]). An optimal base delay in one region may be too short, or too long in another, either causing frequent conflicts or prolonging proposal times unnecessarily.

This variability motivated the development of an adaptive scheduling mechanism that dynamically adjusts delays based on real-time performance metrics. The goal was to optimize both latency and conflict reduction without violating the decentralized leadership principle of CAS Paxos.

### 6.2.2 Simulation Strategy

To analyze the impact of dueling proposers and evaluate our solutions, we built a custom discrete-event simulation framework. This simulator models message timing, network latencies, and consensus attempts, enabling us to replicate realistic distributed system behavior. Because the simulation is discrete-event based, we can compress years of system operation into a manageable timeframe, making it possible to study rare or edge-case scenarios.

We configured the simulator with parameters representative of a production environment and executed a series of experiments, each simulating one hour of operational time. During each experiment, we recorded metrics on both successful and failed proposer rounds as well as overall failure rates.

### 6.2.3 Improved Approach (Adaptive Scheduling and Time-Division Multiplexing)

To address the limitations of a static exponential backoff, we introduced an adaptive strategy that refines backoff intervals based on empirical data. We track the duration of the successful Phase 2 (accept) stage, computed as:

$$D_{\{phase2\}} = T_{\{phase2b_{end}\}} - T_{\{phase2a_{start}\}} \quad (2)$$

where $T_{\{phase2a_{start}\}}$ is the timestamp at the start of Phase 2a and $T_{\{phase2b_{end}\}}$ marks completion of the Phase 2b phase. These durations are added as input to an exponential moving average (EMA) and standard deviation calculation, updated online using Welford's algorithm. We store these statistics in the proposed value itself, ensuring consistency across distributed nodes.

When a proposer receives a NAK, it applies a delay calculated by:

$$\tau_{\{NAK\}} = (\mu_{\{EMA\}} + \sigma) \cdot RandomUniform(0, 2^{\{attempt-1\}}) \quad (3)$$

where $u_{\{EMA\}}$ is the EMA of successful phase 2 durations, $\sigma$ is the standard deviation, and $attempt$ is the retry attempt count. This statistically informed delay allows sufficient time for ongoing Phase 2 operations to conclude, reducing dueling conflicts.

In addition to adaptive backoff, we introduced time-division multiplexing to the Failover Manager's scheduling logic. Rather than adding a random jitter to every proposer's schedule, each proposer references the duration of the most recent successful proposal (excluding conflicts) and shifts its next proposal start by:

$$D_{\{proposal\}} = T_{\{proposal_{end}\}} - T_{\{phase1a_{start}\}} \quad (4)$$

$$\tau_{\{next\}} = T_{\{interval\}} - D_{\{proposal\}} \quad (5)$$

where $D_{\{proposal\}}$ is the observed time from Phase 1 start to proposal completion, and $T_{\{interval\}}$ is the fixed proposer run interval. By adaptively spacing proposals, each proposer reduces its likelihood of colliding with an ongoing round, improving efficiency and success rates.

### 6.2.3 Results

To ensure a comprehensive evaluation, we conducted 10,000 simulations, each representing one hour of operational time. To capture realistic network conditions, we introduced randomly assigned latencies and heterogeneous network characteristics, ensuring a broad range of failure scenarios and proposer contention levels.

The evaluation compared the improved approach, described in the previous section, against our initial implementation. In the initial implementation, conflict handling was based on an exponential backoff with a statically configured base delay, while state update scheduling relied on random jitter to reduce contention. However, as shown in the results, this approach often proved inefficient, as proposers could still initiate conflicting rounds, leading to increased failure rates and prolonged consensus times.

Each simulation included seven acceptors, mirroring production system characteristics. The lease enforcer timeout was set to 45 seconds, and proposers attempted state updates every 30 seconds. To assess system performance under varying contention levels, we tested configurations with 3, 5, 7, and 9 proposers, representing different geographical user regions.

A proposer successfully updates its state and renews its lease at time $T_0$. At $T_1 \approx T_0 + 30s$, it attempts another update. If conflicts prevent completion of Phase 2 of Paxos, the proposer retries. A failure occurs when no successful update is performed within the lease enforcement window, meaning that at $T_2$, where $T_2 - T_0 \geq 45s$, the lease is lost.

While additional system metrics were collected, the primary focus of this evaluation is the failure rate, as it directly reflects the system's ability to mitigate proposer conflicts.

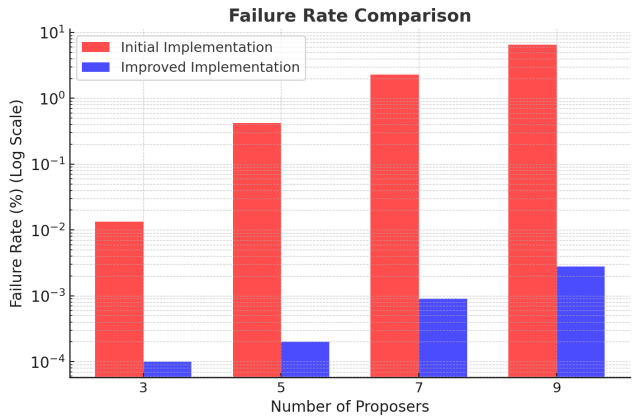

**Figure 9: Failure Rate Reduction improvements with adaptive scheduling and time-division multiplexing**

The failure rate comparison between the initial and improved implementations demonstrates a substantial reduction in proposer conflicts, particularly as the number of proposers increases. In the initial implementation, failure rates increased sharply with higher proposer counts, reaching 6.4950% at nine proposers.

In contrast, the improved approach maintained consistently low failure rates across all test cases, with a maximum of 0.0028% at nine proposers. The adaptive statistics-based conflict handling and structured state update scheduling significantly reduced proposer contention, allowing state updates to complete with minimal interference.

Our findings confirm that the improved approach substantially increases system reliability by reducing proposer conflicts and failure rates. By dynamically managing retries and structuring state updates, the system remains stable even under high contention. These optimizations ensure consistent performance, making this method a viable solution for large-scale distributed environments.

## 7 FUTURE WORK

The Per-Partition Automatic Failover feature is exciting and new, and as with all new features there is more work to do after the initial launch.

### 7.1 Optimizations

Our adaptive backoff strategy currently uses a single, combined exponential moving average calculated for all proposers in a partition-set. We are planning on separating this into using separate per-partition state, where each proposer's statistics are maintained separately. Initial experiments indicate that this strategy further decreases conflicts, especially in scenarios where there are pronounced latency divergence across the regions in a partition-set.

The heartbeating mechanism we implemented to detect failures consumes Cosmos DB resources. It is possible to elide heartbeat updates to the central state when every replica in a partition-set can be assured that all replicas are functioning correctly. This replaces an expensive CAS Paxos heartbeat with a small set of network packets. There is a danger to this optimization: under normal operating conditions, the Failover Manager's acceptor state stores would be under small amounts of load, which would increase when there has been an error. This dual modality behavior introduces risk compared to a model where the load on the acceptor stores stays constant over time.

As discussed earlier, we considered a model using the user's partitions as the state store for the Failover Manager. This model would modestly conserve resources, as most user accounts have a small number of regions and state updates would be limited to these regions. This model also eliminates a dependency on the globally selected regions that back the state stores: as implemented, failure to update a majority of the globally selected regions will prevent the correct operation of a partition-set. These models face significant complexity in the face of adding and removing regions.

For our control plane integration, we took a conservative approach in our first iteration by (mostly) cancelling an in-flight control plane operation if "interrupted" by a failover. Going forward, we will take a deeper look into control plane workflows to identify more scenarios where a control plane operation can be resumed following a failover instead of having to be rolled back. For example, there is no reason in theory for an "Add New Region" operation to be rolled back because we failed over the write-region for one or more partitions.

### 7.2 More external signals

The Failover Manager state machine can admit arbitrary external signals to trigger failovers; we merely need to implement them. Signals indicating a lack of successful traffic from the user's application might be used to trigger failovers. This is common in network outage or misconfiguration scenarios. Integration of application-level monitoring and triggering failover is an area to be explored.

## REFERENCES

[1] D. Rystsov, "CASPaxos: Replicated State Machines without logs," 2018. [Online]. Available: https://arxiv.org/abs/1802.07000.
[2] Microsoft, "Azure Cosmos DB," [Online]. Available: https://learn.microsoft.com/en-us/azure/cosmos-db/. [Accessed 2025].
[3] Microsoft, "Consistency levels in Azure Cosmos DB," 2024. [Online]. Available: https://learn.microsoft.com/en-us/azure/cosmos-db/consistency-levels. [Accessed 2025].
[4] Microsoft, "Partitioning and horizontal scaling in Azure Cosmos DB," 11 2024. [Online]. Available: https://learn.microsoft.com/en-us/azure/cosmos-db/partitioning-overview. [Accessed 2025].
[5] Microsoft, "Commonly asked Service Fabric questions," 2024. [Online]. Available: https://learn.microsoft.com/en-us/azure/service-fabric/service-fabric-common-questions#can-i-create-a-cluster-that-spans-multiple-azure-regions-or-my-own-datacenters. [Accessed 2025].
[6] Microsoft, "Windows Workflow Foundation Programming - Dynamic update," [Online]. Available: https://learn.microsoft.com/en-us/dotnet/framework/windows-workflow-foundation/dynamic-update. [Accessed 2025].
[7] L. Lamport, "The Paxos Algorithm," October 2024. [Online]. Available: https://lamport.azurewebsites.net/tla/paxos-algorithm.html. [Accessed 2025].
[8] T. Grieger, "CASPaxos-tla," [Online]. Available: https://github.com/tbg/caspaxos-tla. [Accessed 2025].
[9] Microsoft, "Transactions and optimistic concurrency control," [Online]. Available: https://learn.microsoft.com/en-us/azure/cosmos-db/nosql/database-transactions-optimistic-concurrency.
[10] Microsoft, "Deploy cloud witness for a failover cluster," February 2025. [Online]. Available: https://learn.microsoft.com/en-us/windows-server/failover-clustering/deploy-cloud-witness. [Accessed 2025].
[11] L. Lamport, "The TLA+ Home Page," [Online]. Available: https://lamport.azurewebsites.net/tla/tla.html.
[12] Microsoft, "Understanding cluster and pool quorum," February 2025. [Online]. Available: https://learn.microsoft.com/en-us/windows-server/storage/storage-spaces/quorum. [Accessed 2025].
[13] D. Lawrence, W. Kumari and P. Sood, "Serving Stale Data to Improve DNS Resiliency," March 2020. [Online]. Available: https://www.rfc-editor.org/info/rfc8767. [Accessed 2025].
[14] L. Lamport, "Paxos made simple.," ACM SIGACT News (Distributed Computing Column), vol. 121, pp. 51-58, December 2001.
[15] Microsoft, "Azure network round-trip latency statistics," [Online]. Available: https://learn.microsoft.com/en-us/azure/networking/azure-network-latency?tabs=Americas%2CWestUS. [Accessed 2025].
[16] Microsoft, "Global distribution with Azure Cosmos DB- under the hood | Microsoft Learn" [Online]. Available Global distribution with Azure Cosmos DB- under the hood | Microsoft Learn. [Accessed 2025].Conference Short Name:WOODSTOCK'18
[17] L. Lamport, D. Malkhi, and L. Zhou. Vertical Paxos and primary-backup replication. Technical report, Microsoft Research, 2009.
[18] L. Lamport, D. Malkhi, and L. Zhou. Reconfiguring a state machine. SIGACT News, 41(1), Mar. 2010.
[19] B. Liskov and J. Cowling. Viewstamped replication revis ited. Technical Report MIT-CSAIL-TR-2012-021, MIT Computer
[20] Iulian Moraru, David G. Andersen, and Michael Kaminsky. 2013. There is more consensus in Egalitarian parliaments. In Proceedings of the Twenty-Fourth ACM Symposium on Operating Systems Principles (SOSP '13). Association for Computing Machinery, New York, NY, USA, 358–372. https://doi.org/10.1145/2517349.2517350
[21] Microsoft, "Consistency level choices - Azure Cosmos DB | Microsoft Learn" [Online]. Available: https://learn.microsoft.com/en-us/azure/cosmos-db/consistency-levels
[22] Lamport, Leslie. Specifying Systems : the TLA+ Language and Tools for Hardware and Software Engineers. Boston :Addison-Wesley, 2003
[23] Fazlali, Mohammad Reza et al. "Raft Consensus Algorithm: an Effective Substitute for Paxos in High Throughput P2P-based Systems." *ArXiv* abs/1911.01231 (2019)
[24] Microsoft, "Azure Cosmos DB SQL SDK connectivity modes," [Online]. Available: https://learn.microsoft.com/en-us/azure/cosmos-db/nosql/sdk-connection-modes
[25] Leslie Lamport. 1998. The part-time parliament. ACM Trans. Comput. Syst. 16, 2 (May 1998), 133–169. https://doi.org/10.1145/279227.279229